\documentclass[copyright,creativecommons,hidelinks]{eptcs}

\usepackage{iftex}
\usepackage{xcolor}

\usepackage{caption}
\usepackage{subcaption}
\usepackage{tikz}
\usetikzlibrary{positioning, shapes.geometric}
\ifpdf
  \usepackage{underscore}         
  \usepackage[T1]{fontenc}        
\else
  \usepackage{breakurl}           
\fi
\usepackage{courier}

\newcommand{\U}{{\ttfamily\symbol{'137}}}
\newcommand{\key}[1]{\texttt{\textbf{#1}}}
\newcommand{\cmt}[1]{\texttt{\textit{#1}}}
\newcommand{\lb}{\texttt{\char`\{}}
\newcommand{\rb}{\texttt{\char`\}}}

\newcounter{lc}  
\DeclareRobustCommand{\showline}{%
  \stepcounter{lc}%
  {\scriptsize\makebox[1mm][r]{\textcolor{blue}{\arabic{lc}}}\hspace{1.5mm}}%
}
\newenvironment{program}{%
  \linespread{1.0}\scriptsize
  \tt\setcounter{lc}{0}
  \begin{tabular}{@{\showline}l@{}}%
  }{%
  \end{tabular}%
}
\newenvironment{programc}{%
  \linespread{1.0}\scriptsize
  \tt
  \begin{tabular}{@{\showline}l@{}}%
  }{%
  \end{tabular}%
}

\title{Specification and Verification for Climate Modeling: Formalization Leading to Impactful Tooling}
\author{Alper Altuntas
\qquad\qquad
Allison H. Baker
\institute{NSF National Center for Atmospheric Research\\ Boulder, CO, USA}
\email{altuntas@ucar.edu \qquad\qquad abaker@ucar.edu}
\and
John Baugh
\institute{North Carolina State University\\
Raleigh, NC, USA}
\email{jwb@ncsu.edu}
\and
Ganesh Gopalakrishnan
\institute{University of Utah\\
Salt Lake City, UT, USA}
\email{ganesh@cs.utah.edu}
\and
Stephen F. Siegel
\institute{University of Delaware\\
Newark, DE, USA}
\email{siegel@udel.edu}
}
\def\titlerunning{Specification and Verification for Climate Modeling}
\def\authorrunning{Altuntas, Baker, Baugh, Gopalakrishnan, and Siegel}
\begin{document}
\maketitle

\begin{abstract}

Earth System Models (ESMs) are critical for understanding past climates and projecting future scenarios. However, the complexity of these models, which include large code bases, a wide community of developers, and diverse computational platforms, poses significant challenges for software quality assurance. The increasing adoption of GPUs and heterogeneous architectures further complicates verification efforts. Traditional verification methods often rely on bitwise reproducibility, which is not always feasible, particularly under new compilers or hardware. Manual expert evaluation, on the other hand, is subjective and time-consuming. Formal methods offer a mathematically rigorous alternative, yet their application in ESM development has been limited due to the lack of climate model-specific representations and tools. Here, we advocate for the broader adoption of formal methods in climate modeling. In particular, we identify key aspects of ESMs that are well suited to formal specification and introduce abstraction approaches for a tailored framework. To demonstrate this approach, we present a case study using CIVL model checker to formally verify a bug fix in an ocean mixing parameterization scheme. Our goal is to develop accessible, domain-specific formal tools that enhance model confidence and support more efficient and reliable ESM development.

\end{abstract}

\section{Introduction}

Increasingly frequent extreme weather events~\cite{IPCC2021} and possible climate tipping points such as the weakening or collapse of the Atlantic Meridional Overturning Circulation~\cite{boers2021observation} pose significant risks. Earth system models (ESMs) are among the most comprehensive tools for studying such phenomena by capturing the complexity of climate dynamics and combining atmospheric, oceanic, land, and cryospheric processes into a coupled framework. Over decades, these models have evolved to encompass millions of lines of code that represent a wide range of physical, chemical, and biological processes governing the Earth system. Through the representation of such diverse and interacting processes, ESMs enable researchers to explore complex feedback mechanisms, predict potential future scenarios, and inform both our understanding and policy decisions.

\begin{figure}
    \centering
    \includegraphics[width=0.5\textwidth]{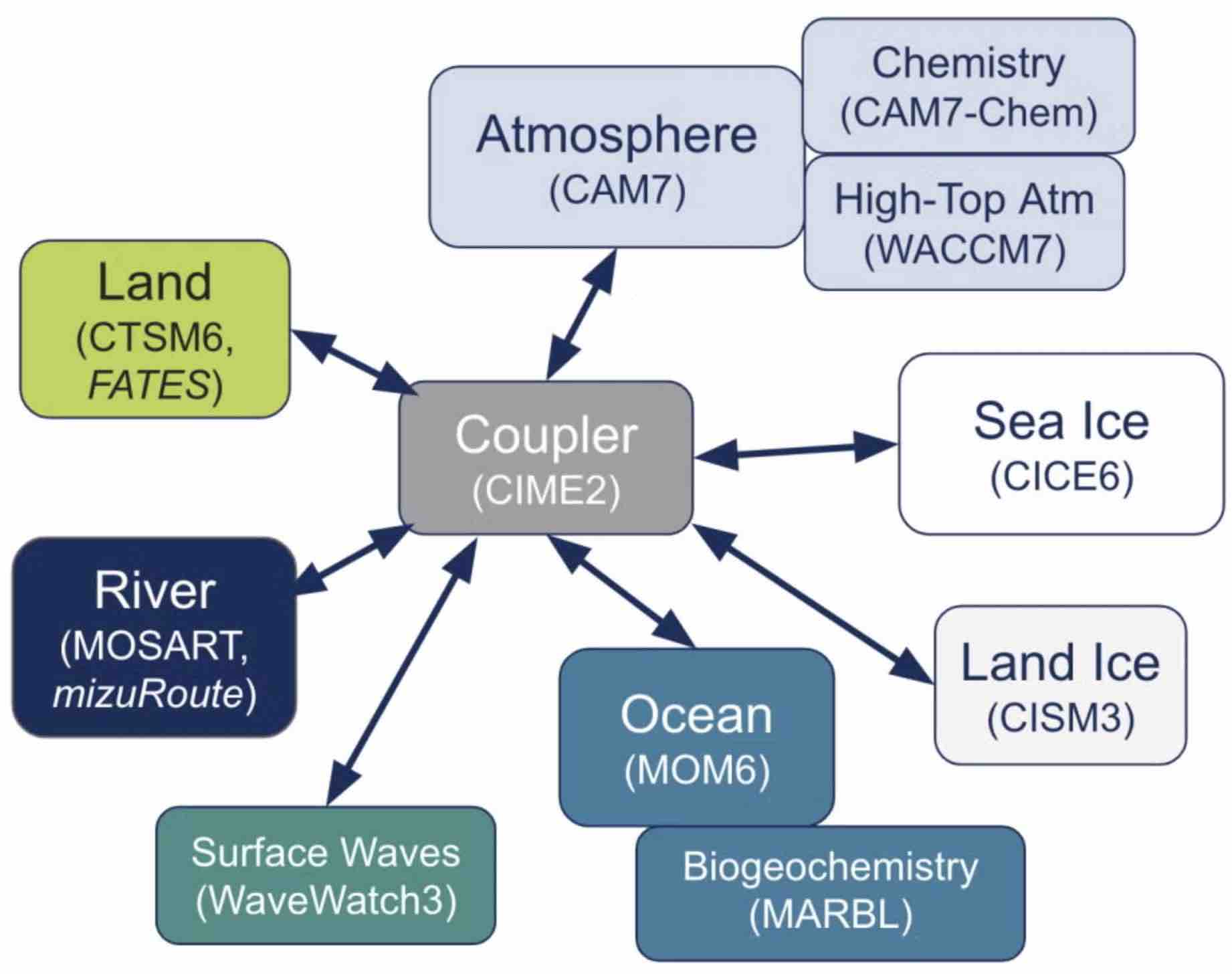}
    \caption{The components of the latest version of the Community Earth System Model (CESM3). Each component interacts with the rest of the system via the coupler in a hub-and-spoke structure~\cite{lawrence2024cesm3}.}
    \label{fig:cesm}
\end{figure}

To accurately simulate the behavior of each subsystem, ESMs are composed of a number of components, each responsible for computing the time evolution of physical phenomena within their respective domains. For example, atmospheric components simulate wind patterns, temperature, and precipitation dynamics, while ocean components model circulation, heat transport, and mixing. The coupled system typically includes a driver, which orchestrates execution during runtime, and a coupler, which handles the exchange of data between components. Figure~\ref{fig:cesm} illustrates the components of the latest version of a widely used model, the Community Earth System Model (CESM), and its hub-and-spoke architecture.

Because of their inherent complexity and near continuous development, ESMs are extremely challenging to maintain and validate. Software quality assurance for ESMs relies heavily on traditional testing methodologies, including regression tests that often depend on bitwise reproducibility, meaning that a simulation must produce identical numerical results down to the last bit when re-run with the same inputs. When exact reproducibility is not feasible, such as under new compilers or hardware, field experts must manually evaluate model outcomes, which is both time-consuming and subjective \cite{Altuntas2024}. While ensemble consistency testing~\cite{baker2015new, milroy2018} offers solutions in such scenarios, development of new modules or algorithms, especially those lacking existing oracles, necessitates additional verification methods.

In this paper, we share our perspectives on how formal methods can be utilized for specifying and verifying ESM processes and mechanisms. Towards this goal, we explore various abstractions and modeling approaches to facilitate such specification and verification efforts. Since formal specification languages and tools often demand specialized expertise and lack the necessary constructs to effectively represent ESM-specific concepts, we envision a comprehensive framework that incorporates ESM-specific constructs, including the continuous evolution of quantities, discrete updates, spatial discretizations, and concurrency. By introducing abstractions tailored to these features, we aim to lower the barriers to applying formal methods, providing explicit representations that facilitate reasoning about data dependencies and algorithmic machinery found in ESMs. Consequently, our approach will enable the formal verification of properties of interest that are typically tested implicitly or through ad hoc methods, thereby enhancing the reliability and correctness of ESMs. This can complement existing verification and validation efforts for both the development of new models and the analysis of existing codebases.

This paper is organized as follows. In Section 2, we identify the key modeling aspects in ESM development that are amenable to formal specification and highlight the challenges arising from these aspects. Section 3 provides a summary of current practices in ESM verification and validation. In Section 4, we discuss our perspectives and abstraction approaches towards a new specification framework. Section 5 demonstrates these concepts through a test case, illustrating how the proposed abstractions can be applied in practice. Finally, Section 6 offers concluding remarks and future directions of this work.

\section{Targeted Modeling Aspects}

Here, we summarize some high-level modeling aspects and their associated challenges, which are critical for ensuring correctness and are well-suited to the application of formal specification and verification.

\subsection{Model Coupling}

At every coupling timestep, which is typically on the order of an hour, states and fluxes are exchanged along the interfaces between ESM components that interact. This process, known as inter-component coupling, is the mechanism by which the components inform each other and incorporate the effects of other subsystems. For instance, the ocean component relies on inputs such as atmospheric pressure, wind stress, and heat flux to compute surface currents, temperature changes, and other ocean dynamics. Similarly, the land component requires precipitation and radiation inputs from the atmosphere to simulate processes like runoff and energy exchange. Such data dependencies pose constraints that necessitate careful configuration of run sequencing and processor layout.

In addition to inter-component coupling, ESMs also employ intra-component coupling mechanisms, such as one-way and two-way nesting and adaptive mesh refinement. Furthermore, the combination of inter- and intra-component coupling, such as those found in multi-instance runs~\cite{altuntas2017verifying}, introduces additional complexities. The increasing use of Machine Learning (ML), whether for parameterizations or as complete surrogate models, further complicates module dependencies, system robustness, and the handling of edge cases. 


\subsection{Continuity and Discreteness}

A key aspect of our proposed specification approach is the distinction and the interplay between continuous and instantaneous processes. This distinction aligns with the breakdown of ESM components into three main categories: (1) dynamical core, (2) parameterizations, and (3) computational methods.

\begin{figure}
        \centering
        \includegraphics[width=1.0\textwidth]{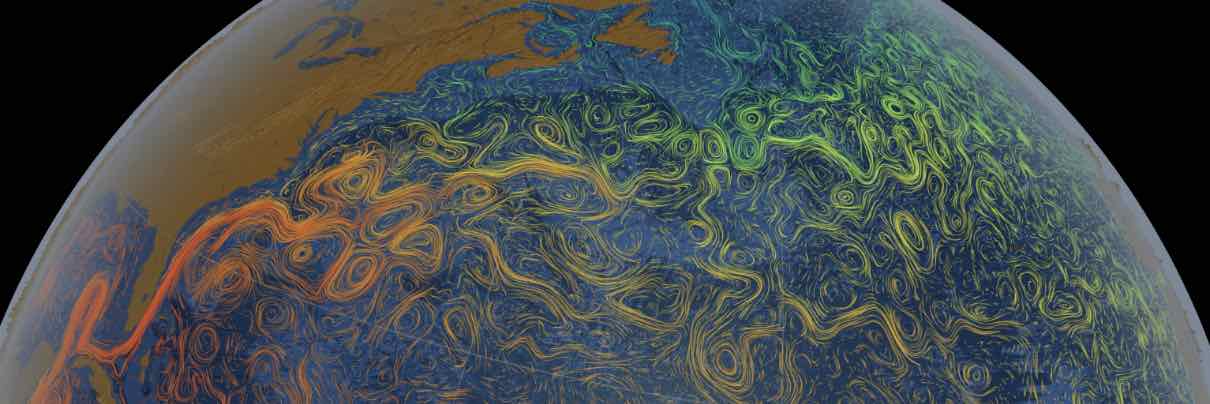}
        \caption{The visualization of the Gulf Stream as simulated by the MIT General Circulation Model (MITgcm). The ocean flows are color-coded with sea surface temperature data. Adapted from \cite{menemenlis2008ecco2, nasasvsgulfstream}.}
        \label{fig:current}
    \end{figure}

The governing fluid and thermodynamic equations, formulated and solved numerically as partial differential equations (PDEs), form the foundation of \textbf{\emph{dynamical cores}} \cite{thuburn2008some}. These equations describe the resolved large-scale motion of fluids, such as air in the atmosphere or water in the ocean, as well as the associated energy and mass transport processes. Although these PDEs are discretized in time and space to enable numerical computation, they are fundamentally continuous in nature. That is, they describe smooth changes in the state variables (e.g., velocity, temperature, and density) over time and space. This smoothness contrasts with the inherently discontinuous or sub-grid-scale processes, which are represented differently in ESMs. As an example of a continuous process, Figure~\ref{fig:current} illustrates the smooth evolution of ocean currents simulated by a model~\cite{menemenlis2008ecco2, nasasvsgulfstream}.

\textbf{\emph{Parameterizations}}, on the other hand, encapsulate empirical and theoretical algorithms that represent processes unresolved by the model’s spatial and temporal resolution. Examples include turbulence, cloud formation, and ocean mixing. Some of these processes may appear \emph{instantaneous}, for instance, when triggered by threshold-based or switch-like criteria, as with wetting and drying algorithms in ocean models~\cite{baugh2018formal} or ice-sheet calving~\cite{benn2007calving}. Others, such as vertical ocean mixing~\cite{large1994oceanic} and radiative transfer~\cite{pincus2013paths}, generally evolve more smoothly over time. Even in schemes considered ``continuous,'' discontinuities can still occur. For example, in the K-Profile Parameterization (KPP)~\cite{large1994oceanic}, the ocean boundary-layer depth is determined by stability criteria that may shift from one timestep to the next, causing changes in mixing rates. Parameterizations are critical for closing the system of equations and ensuring that models remain computationally feasible while capturing the essential behavior of unresolved processes. However, their reliance on empirical approximations introduces uncertainties that affect the overall reliability of model predictions. Unlike the governing equations in dynamical cores, parameterizations often lack direct theoretical or analytical ``oracles'' to guide their design, and they are often algorithmically complex with extensive logical branching (e.g., the KPP scheme~\cite{large1994oceanic}). This complexity is compounded by the interplay between continuous and instantaneous processes, which adds significant challenges to accurately representing these sub-grid-scale phenomena.

Beyond the physical processes described above, ESMs also employ a range of \textbf{\emph{computational methods}} to ensure numerical accuracy, stability, and efficiency. Examples include remapping in Arbitrary Lagrangian–Eulerian (ALE) algorithms~\cite{griffies2020primer}, adaptive mesh refinement~\cite{zhang2019amrex}, and halo exchanges, where parallel communication is used to exchange boundary data between adjacent computational subdomains assigned to each processing element. These steps, typically executed \emph{instantaneously} at predefined points in the simulation workflow (e.g., after each timestep), do not represent physical phenomena but are instead essential for preserving solution integrity across evolving model grids, coordinating parallel processes, and maintaining computational performance.


\subsection{Spatial Discretization}

The governing equations in a dynamical core must be discretized in both time and space to enable numerical solutions. In the spatial dimension, this involves using a grid to represent either the entire globe (in global models) or a selected region (in regional models).  ESM grids are categorized as either structured or unstructured. Structured grids exhibit uniform connectivity, whereas unstructured grids accommodate irregular connectivity and cell shapes.

\begin{figure}
\centering
\begin{subfigure}{.65\textwidth}
  \centering
  \includegraphics[width=\linewidth]{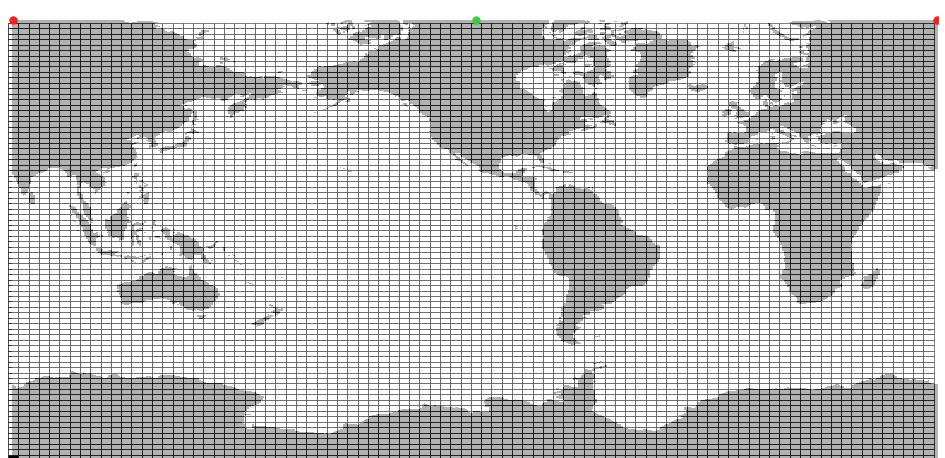}
  \caption{logical index space}
  \label{fig:sub1}
\end{subfigure}%
\quad
\begin{subfigure}{.3\textwidth}
  \centering
  \includegraphics[width=\linewidth]{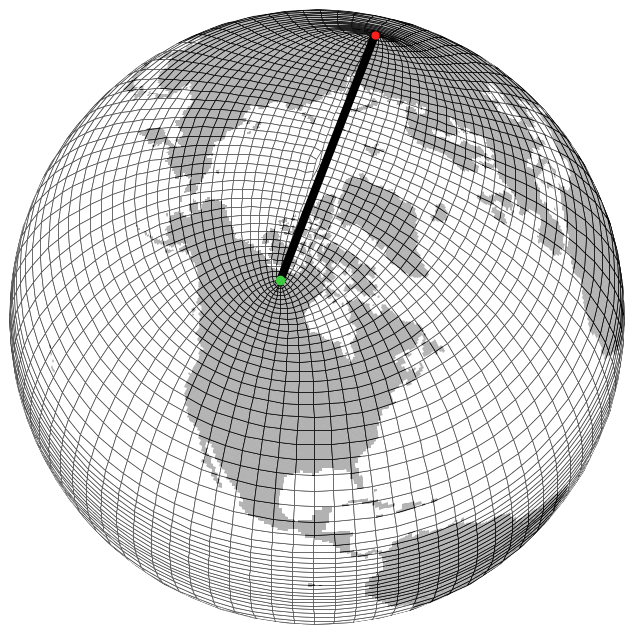}
  \caption{geographical}
  \label{fig:sub2}
\end{subfigure}
    \caption{A tripolar grid. These are commonly used in ocean models to avoid singularities in open water by shifting the nominal pole onto land. Panel~(a) illustrates the grid in index space, where the northern edge is periodically folded to create the tripolar stitch along the grid lines connecting the two red points (same geographical location) to the green point. Panel~(b) displays the same grid in geographical coordinates, with the dark line between the singularity points indicating the tripolar stitch. 
    }

        \label{fig:t232}\label{fig:test}
\end{figure}

Here, we focus on structured, logically rectangular grids, among the most common discretization choices in ESMs. These grids preserve a consistent indexing scheme in each dimension, even when mapped onto spherical geometries and curvilinear coordinates, thereby simplifying numerical methods. Figure~\ref{fig:t232} shows an example of a structured grid, specifically a tripolar grid, frequently used in ocean modeling to avoid complications arising from polar singularities in latitude–longitude grids.

The representation of spatial grids in ESMs presents a set of challenges. Each component typically operates on a unique grid tailored to its resolution and physical characteristics, requiring remapping or interpolation for data exchange across components with different grids. The placement or “staggering” of physical and dynamical fields within grid cells must be carefully handled to avoid numerical errors and ensure consistency. Additionally, periodic boundary conditions (e.g., wrapping east–west boundaries or folding the north edge in tripolar grids) must be correctly implemented to maintain continuity. Finally, adaptive approaches, such as adaptive mesh refinement or Arbitrary Lagrangian–Eulerian (ALE) algorithms introduce further challenges. 

The decomposition of model grids for parallel computing introduces even greater complexity. The model domain is partitioned into subdomains, each assigned to a processor, with halo exchanges occurring at subdomain boundaries to facilitate interprocessor communication through consistent variable updates.  The shift toward GPU-based systems and hybrid parallelism (e.g., combining MPI with OpenMP or GPU directives) adds new demands for code refactoring, data transfer optimization, and careful management of mixed CPU–GPU resources. 


\section{Current verification and validation practices}


Here, we briefly describe current and previous approaches to evaluating software correctness in the climate modeling community. Specifically, we summarize scientific methods based on domain expertise, traditional software techniques such as testing, and novel approaches leveraging formal verification.

\subsection{Scientific Methods}

Climate scientists often evaluate code changes in a time-consuming and subjective manner that involves comparing new experimental results with control runs and observational data \cite{easterbrook2009}.  For example, hindcasting is a common technique that involves running a model simulation from a past state with known conditions (e.g.,  pre-industrial) through to the present day.  The quality of the model output can then be evaluated against available observational data from the recent past.  While this approach has been used since the early days of climate modeling, it requires extensive computational resources for the long simulations, and exhaustively testing the many available model configurations (e.g., grid type, resolution, initial and boundary conditions, numerical method, etc.), is not practical \cite{clune2011}. Further, the manner in which domain experts analyze and compare model data is not standardized and typically dependent on the scientist's interests and expertise.  Another means of evaluating scientific correctness of models is via benchmarks and model intercomparisons.  Benchmarks have long been popular tools in the high-performance computing community, and have recently gained traction for evaluating correctness and performance in climate modeling. As an example, the High Performance Climate and Weather (HPCW) benchmark \cite{hpcw} comprises a collection of realistic, near-operational weather and climate workloads, encompasses a number of popular models, and incorporates test cases, verification procedures, and metrics for comparative analysis. Model intercomparison projects are also quite popular and necessary for assessing the relative strengths and weaknesses of various coupled climate models through detailed comparisons. Most notably, modeling centers around the globe participate in the Coupled Model Intercomparison Projects (CMIPs) \cite{cmip}.  Each CMIP, including most recent CMIP phase 6 (CMIP6) \cite{cmip6}, contains a variety of experiments designed to evaluate past, present, and future climate states.  The CMIP results are particularly useful for quantifying the uncertainty for future climate projections, and  understanding factors in differences across model projections leads to improvements in future model development.


\subsection{Software Methods}

Software testing to verify model behavior and detect bugs is critical when developing large and complex simulation codes like ESMs.  
While climate modeling centers regularly employ various traditional testing approaches such as unit testing, regression testing, and system testing, these approaches should be used to a much larger extent than currently done \cite{Altuntas2024}. 
Though unfortunately, many tools and technologies developed thus far are not yet suited for large and complex HPC codes like ESMs \cite{gokhale2023report}, and this issue leads to the frequent utilization of application-specific custom tests in the climate modeling community to verify model-specific behaviors.  
We note that a particular challenge for verification of these models is that yielding bitwise reproducible results is often unattainable due to the chaotic nature of these models and the diverse hardware and software environments in which they operate.  
When bitwise reproducibility is not possible, scientists may be faced with the challenge of determining whether the difference in simulation results is due to an error or simply to the model’s natural variability.
And in this case, experts may be more likely to evaluate model results in a time-consuming and subjective manner.

Addressing this correctness question without bitwise reproducibility is somewhat tractable when the difference (e.g., a compiler upgrade, a port or optimization for different hardware, or lack of associativity in floating-point arithmetic) is not expected to change the scientific conclusions, and in such cases a statistical approach can be utilized.  
In particular, rather than determine whether new results from a non bit-reproducible modification are correct, we address whether the new results are statistically distinguishable from the previous (control) results. 
An example of such a statistical approach is a technique known as ensemble consistency testing, which is being increasingly utilized (e.g., \cite{baker2015new, milroy2018}). 

Yet many modifications in the software lifecycle, such as adding a new physical process, are expected to produce “answer-changing” results. Moreover, testing is inherently incomplete; as Dijkstra famously stated, “Testing can be used to show the presence of bugs, but never to show their absence”~\cite{dijkstra1970notes}, further underscoring the need for complementary verification techniques beyond traditional testing.

        
\subsection{Formal Specification and Verification}\label{pastformalspec}


Formal methods~\cite{clarke96, wing90, woodcock09} are mathematically based techniques for specifying, analyzing, and verifying system properties throughout the development lifecycle. They provide a systematic alternative to ad hoc approaches, promoting correctness, consistency, and completeness. At the specification level, they enable precise descriptions of system behavior without constraining implementation. In design, techniques like data refinement and simulation proofs support structured development. At the implementation level, methods such as Floyd-Hoare logic ensure correctness by using logical assertions to establish that a program meets its specification. Approaches range from theorem proving and model checking to automatic code generation.

Existing tools aim to balance expressiveness and tractability, often employing semantic restrictions to make models easier to reason about and ensure computational feasibility. These tools are typically tailored for specific domains, such as hardware verification, communication protocols, or safety-critical systems, and are optimized to verify the kinds of correctness properties most relevant to those contexts---such as deadlock freedom, synchronization, or logical consistency.

Adapting them for scientific computing introduces additional challenges. Not only must they handle the inherent complexity of large-scale systems, but they must also abstract from or integrate aspects of numerical analysis, such as stability and precision, in their frameworks. Moreover, identifying and targeting bugs most relevant to scientific computing---whether they arise from software structure, parallel execution, or numerical interactions---requires rethinking how existing tools can be extended or repurposed for this domain.

Recent work in applying formal methods to scientific computing has developed along two complementary lines. The first incorporates numerical concerns directly into the reasoning process. Bientinesi et al.~\cite{bientinesi2005} use Floyd-Hoare logic to derive and prove the correctness of dense linear algebra algorithms, employing simplified matrix notation to hide indexing details and systematically identify invariants for nested loops.
Kellison et al.~\cite{kellison-etal:2023:laproof} used deductive verification techniques, rooted in the Coq proof assistant \cite{coq-web} and the Verified Software Toolchain, to prove accuracy and correctness claims of several basic linear algebra routines expressed in C, including sparse matrix algorithms.
Siegel et al.~\cite{siegel2008} present a framework for verifying real, IEEE floating-point, or exact equivalence between sequential and parallel numerical programs, using symbolic expressions and the SPIN model checker.  
The CIVL Model Checker \cite{siegel-etal:2015:civlsc} takes a similar approach to the analysis of scientific C code that uses MPI, OpenMP, CUDA, or Pthreads.
Symbolic Taylor Forms---a rigorous methods for tightly bounding
floating-point errors---was introduced in~\cite{solovyev-toplas-2019} and significantly scaled up in~\cite{satire-sc20-das}, with conditional branch instability analyzed in~\cite{seesaw-cluster-21-das}.
Mapping to heterogeneous hardware comes with many additional defect-possibilities that are surveyed in~\cite{rising-heterogeneity-comport-gang} with a GPU-specific tool for detecting floating-point exceptions contributed by~\cite{hpdc2023-gpu-fpx}.

Altuntas and Baugh~\cite{altuntas2018hybrid} apply hybrid theorem proving to model numerical software as hybrid systems, combining discrete updates and continuous dynamics. Their case study identifies undesirable behavior encountered in ocean components of climate models, demonstrating the potential of this approach to improve confidence in numerical simulations.

The second line separates concerns by abstracting from numerical computations and focusing on the software structure that integrates and manages these computations, emphasizing structural correctness in a variety of applications. Baugh et al.~\cite{baugh2018formal} use Alloy to verify equivalence-preserving extensions to ADCIRC, a large-scale hurricane storm surge model used in production, ensuring the soundness of implementation choices. Dyer et al.~\cite{dyer-correctness-2019} demonstrate how sparse matrix operations, including multiplication and format conversions, can be modeled and verified using Alloy, relying on bounded analysis to ensure structural refinement. Benavides et al.~\cite{benavides-lcpc-2023} explore refinement checking in high-performance computing applications, using Alloy to model single program, multiple data (SPMD) parallelism in Coarray Fortran, with an example of an iterative Jacobi solver for Laplace’s equation.


\section{Towards a Relational Spatio-Temporal Logic}

In this section, we introduce our abstraction approaches and perspectives for effectively specifying key aspects of ESMs. Our aim is to establish a foundation for a relational spatio-temporal logic, i.e., a modeling framework that enables ESM developers to specify critical structures and behaviors of interest within ESMs, reason about their interactions, and formally verify key correctness properties.

\subsection{Model Coupling and Concurrency}

Concurrency in Earth System Models (ESMs) occurs across multiple levels: from inter-component and inter-process interactions to GPU thread management. These instances of concurrency within ESMs present unique challenges but also offer opportunities for formal specification, particularly due to the modular structure of most ESMs, where concurrency typically occurs between well-defined interfaces. For example, at the intercomponent level, the thin "model cap" layers found in popular models like CESM, along with the coupler, provide natural boundaries for defining assumptions and guarantees.

As such, we propose representing concurrency via compositional reasoning, e.g., assume-guarantee reasoning~\cite{Flanagan2003,Hayes2018}. This methodology has been successfully applied in other contexts~\cite{fan2017dryvr, rajagopal2023using}, and is similarly well-suited for our purposes since explicitly specifying every component of a coupled ESM is impractical. Instead, a targeted strategy similar to lightweight formal methods is both necessary and achievable. Within our specification framework, we therefore propose a higher-level analysis and representation that emphasizes data dependencies and the sequencing of events.

A practical example of this approach can be utilized in reasoning about the run sequencing of model components within CESM and similar models. Until recently, run sequencing was implemented through hard-coded subroutine calls. More recently, configuration input files, known as NUOPC run sequencing files~\cite{nuopcwebsite}, have been used to specify run sequencing in a more configurable manner, marking significant progress. However, there is still room for improvement: the rationale behind certain run sequencing configurations is sometimes unclear. By explicitly defining dependencies rather than merely specifying the resulting sequence, run sequencing can be automatically generated using techniques such as SAT solvers, thereby facilitating the optimization of concurrency, all while preserving the integrity of the scientific results.

As for process-level concurrency, which is extensively studied in the formal methods community, we propose leveraging existing tools such as CIVL to establish equivalence between sequential and parallel numerical programs~\cite{siegel2008}. Similarly, for GPU parallelism, tools such as Faial~\cite{faial-oopsla-2024} and HiRace~\cite{jacobson2024hirace} can be employed to detect  data races, thus helping enhance correctness at the thread level.

\subsection{Continuous and Instantaneous Processes}

Our approach to effectively specifying and reasoning about continuous and instantaneous behaviors draws inspiration from methodologies in the cyber-physical systems (CPS) domain. In CPS, modelers abstract the inherently continuous real world by treating certain rapidly occurring events as instantaneous, e.g., modeling the bouncing of a ball~\cite{platzer2018logical}.

In contrast, numerical modeling software, such as Earth System Models (ESMs), are fundamentally discrete. However, some processes, such as those simulated by dynamical cores, evolve smoothly over time. We therefore adopt a similar abstraction in the opposite direction: by assuming continuity for these smoothly evolving processes, we can focus on their interactions and facilitate efficient reasoning. Instead of attempting to incorporate the discrete computation of partial differential equations (PDEs) directly, which becomes intractable with even a small number of timesteps and grid points, we propose modeling PDEs as continuous in time but discrete in space. This abstraction, commonly used in numerical analysis, is known as the method of lines, where a PDE system is transformed into a system of ordinary differential equations (ODEs). By doing so, we enable the use of formal verification techniques designed for continuous systems, such as those developed in the CPS community.

We list several alternative methods for specifying PDEs as ODEs and reasoning about their behavior: One such method is \emph{Hybrid Programming} along with \emph{Differential Dynamic Logic} \cite{platzer2018logical}, which allows for the specification of differential invariants to reason about continuous behavior without explicitly solving the ODE system, as demonstrated in \cite{altuntas2018hybrid}. Another possible approach is \emph{symbolic solution} of ODEs, particularly when closed-form solutions exist. In cases where closed-form solutions are unavailable, \emph{numerical solutions} of ODEs, as is done in the case study in the next section, may be employed. Additionally, to increase confidence in numerical solutions, \emph{probabilistic and stochastic approaches} can be used. These include numerically computing ODEs multiple times with precision-level perturbations and running multiple ensembles, inspired by ensemble consistency testing \cite{baker2015new}, or employing stochastic methods commonly used in the CPS domain for reachability analysis \cite{bulychev2012uppaal}. Finally, when dealing with original PDEs that are too complex for direct integration into a formal framework, simpler \emph{surrogate ODEs} (e.g., linear shallow water equations instead of nonlinear momentum and continuity equations) or \emph{nondeterministic representations} of continuous behavior can be utilized.

By employing one or a combination of these methods, we can effectively model the continuous time evolution of ESMs while maintaining computational tractability and facilitating formal reasoning about interactions and system behavior. 

\paragraph{Reals vs. Floating-Point Arithmetic} Floating-point arithmetic introduces a range of issues that can compromise model correctness, including rounding errors, precision limitations, and numerical instability. However, maintaining continuity in time often necessitates working with real numbers rather than floating-point representations. This requirement is particularly relevant for methods such as differential dynamic logic and symbolic solutions. While reasoning about floating-point arithmetic is crucial, especially when feasible, it often introduces significant complexity and computational overhead that can impede formal verification efforts. Although the issues related to floating-point arithmetic are critical and remain active areas of research, by abstracting away from the intricacies of floating-point computations, we can focus on higher-level aspects of the model, such as the correctness of algorithms, data dependencies, and the formal verification of interactions between components. This abstraction enables a clearer and more manageable framework for reasoning about system behavior, facilitating the identification and resolution of logical and structural issues that are often overshadowed by numerical concerns.

\paragraph{Instantaneous events:} Specifying discrete algorithms, such as parameterization routines and computational processes, is relatively straightforward when integrating them into an abstract model. This is because they map naturally onto discrete transitions, which are the core building blocks of most formal methods. However, there is a crucial distinction: ESM developers typically utilize procedural programming paradigms, whereas formal specification relies on declarative paradigms. Bridging this gap is essential. Towards that end, one approach may be to adopt intermediate representations, for example, by using state transition operators as implemented in tools like Alloy 6.

\subsection{Mesh representations}

In ESM codebases, grids are typically represented as \emph{structures of arrays}, where each horizontal grid property (e.g., the coordinates, areas, angles, etc.) is stored in two-dimensional arrays. This also applies to the physical quantities incident on grid cells (e.g., temperature, density, fluxes, etc.). The indices 
\texttt{i} and \texttt{j}, then, correspond to the logical row and column identifiers that locate each cell (or node) in the global model domain, abstracting away the underlying spherical or regional map projection.

In our abstraction approach, we propose maintaining this logical indexing-based representation. However, we argue for the specification of a minimal grid: as small as the local domain of dependence~\cite{altuntas2018hybrid}. In practice, this often means focusing on a single grid cell (or node) and its immediate neighbors in each dimension, sufficient to capture the local PDE stencil or parameterization dependencies. This is justified because both discrete and continuous processes generally have a very limited domain of dependence. In the case of continuous processes, this is particularly true since hyperbolic PDEs (e.g., those governing fluid flow) have a finite propagation speed. The Courant-Friedrichs-Lewy (CFL) condition restricts how far information can travel in a single timestep, ensuring that any local update only depends on a limited set of neighboring cells. Consequently, spatial data dependencies do not extend arbitrarily beyond those nearby cells, so abstracting the global domain into a small grid region plus nondeterministic external inputs allows compositional reasoning without enumerating the entire grid.

Representing staggering, i.e., the placement of quantities on different locations of a grid cell (e.g., cell center, edges, or corners), can be based on viewing grids as a Hasse diagram, as was done by Knepley et al. for parallel mesh and data distribution, load balancing, and overlap generation~\cite{knepley2015unstructured}, or similarly via relational logic~\cite{baugh2018formal}. These approaches facilitate reasoning about adaptive algorithms like AMR, parallel operations such as halo exchanges, east-west wrapping, and other specialized boundary handling mechanisms that maintain grid continuity, as well as other grid-related algorithms essential for ensuring computational accuracy and efficiency.

\section{A Case study}

The hybrid model studied in \cite{altuntas2018hybrid} is known as the K-profile parameterization (KPP) scheme. This model, which is available online \cite{altuntas2018keymaera}, involves control iterating a finite, though unspecified, number of times.  At each iteration, there is a sequence of discrete, instantaneous updates to the state, followed by the continuous evolution of the state for some non-zero, finite amount of time.  The continuous evolution is specified by an ordinary differential equation.   In this case, the equation has the form $y'=-y$, where $y'$ is the time-derivative of variable $y$.  KeYmaera X~\cite{fulton2015keymaera} was used to find a defect in this scheme in which a variable $K$, which should always be positive, in fact becomes negative.  The defect was repaired by modifying the function that updates $K$, and KeYmaera was used to construct a proof, in differential dynamic logic, that $K$ remains positive and other invariants hold. 

\newcounter{innerstart}
\newcounter{innerend}
\newcounter{assertK}

\begin{figure}[t]
  \begin{tabular}{@{}ll@{}}
    \begin{minipage}[t]{.60\columnwidth}\small\vspace{0pt}
\begin{program}
\key{\$input} \key{int} N; \cmt{// number of outer iterations}\\
\key{\$input} \key{int} M; \cmt{// max number inner iterations}\\
\key{\$input} \key{double} dt; \cmt{// delta\_t}\\
\key{\$assume}(0<dt \&\& dt<1);\\
\key{\$input} \key{double} zw, D, w;\\
\key{\$assume}(D>0 \&\& D>zw \&\& zw>0 \&\& w>0);\\
\key{double} t=0.0; \cmt{// time} \\
\key{double} nu, dnu, h, sigma, alpha, zCr, K, a2, a3;\\
\\
\key{double} G(\key{double} sigma, \key{double} a2, \key{double} a3) \lb\\
\ \ \key{return} sigma + \\
\ \ \ \ a2*\key{\$pow}(sigma,2) + a3*\key{\$pow}(sigma,3);\\
\rb\\
\\
\key{void} computeNu(\key{void}) \lb\\
\ \ \key{\$havoc}(\&nu);\\
\ \ \key{\$assume}(nu>0);\\
\ \ \key{\$havoc}(\&dnu);\\
\rb\\
\\
\key{void} computeBLD(\key{void}) \lb\\
\ \ h = D - zCr;\\
\ \ sigma = (D - zw)/h;\\
\ \ \key{\$havoc}(\&alpha);\\
\ \ \key{\$assume}(0<alpha \&\& alpha<1);\\
\ \ zCr = alpha*zCr;\\
\rb\\
\\
\key{void} computeK(\key{void}) \lb\\
\ \ a2 = -2 + 3*nu/(h*w) + dnu/w;\\
\ \ a3 = 1 - nu/(h*w) - dnu/w;\\
\ \ K = h*w * G(sigma, a2, a3);\\
\rb\\
\\
\key{void} invariant(\key{void}) \lb\\
\setcounter{assertK}{\value{lc}}%
\ \ \key{\$assert}(K>0);\\
\ \ \key{\$assert}(zw>=zCr);\\
\ \ \key{\$assert}(zCr>0);\\
\rb
\end{program}
    \end{minipage}
    &
    \begin{minipage}[t]{.35\columnwidth}\small\vspace{0pt}
\begin{programc}
\key{void} initialConditions() \lb\\
\ \ \key{\$havoc}(\&nu);\\
\ \ \key{\$havoc}(\&dnu);\\
\ \ \key{\$havoc}(\&h);\\
\ \ \key{\$havoc}(\&sigma);\\
\ \ \key{\$havoc}(\&alpha);\\
\ \ \key{\$havoc}(\&zCr);\\
\ \ \key{\$havoc}(\&K);\\
\ \ \key{\$havoc}(\&a2);\\
\ \ \key{\$havoc}(\&a3);\\
\ \ \key{\$assume}(0<nu \&\& K>0);\\
\ \ \key{\$assume}(zCr == zw);\\
\rb\\
\\
\key{void} printState(\key{void}) \lb\ ... \rb\\
\\
\key{int} main(\key{void}) \lb\\
\ \ printState();\\
\ \ initialConditions();\\
\ \ \key{for} (\key{int} i=0; i<N; i++) \lb\\
\ \ \ \ printState();\\
\ \ \ \ invariant();\\
\ \ \ \ computeBLD();\\
\ \ \ \ computeNu();\\
\ \ \ \ computeK();\\
\ \ \ \ \cmt{// zCr'=-zCr ...}\\
\setcounter{innerstart}{\value{lc}}%
\ \ \ \ \key{int} m = \key{\$choose\U{}int}(M);\\
\ \ \ \ \key{for} (\key{int} j=0; j<m; j++) \lb\\
\ \ \ \ \ \ t += dt;\\
\ \ \ \ \ \ zCr += -zCr*dt;\\
\setcounter{innerend}{\value{lc}}%
\ \ \ \ \rb\\
\ \ \rb\\
\ \ printState();\\
\ \ invariant();\\
\rb
\end{programc}
    \end{minipage}
  \end{tabular}
  \caption{CIVL-C model of the KPP defect fix.}
  \label{fig:civl-kpp}
\end{figure}

Here we describe a more lightweight modeling approach for the same problem.  We wrote a model of the hybrid program in CIVL-C, the language of the CIVL Model Checker (Fig.\ \ref{fig:civl-kpp}).  CIVL-C is an intermediate verification language with syntax and semantics that are similar to other imperative programming languages.  The model checker analyzes a CIVL-C model using symbolic execution, i.e., the values assigned to variables are symbolic expressions rather than concrete numbers.

There is no way to express the continuous evolution of a quantity, governed by an ODE, in CIVL-C.  Instead, we have modeled this by a simple first order scheme to approximate the result.  This involves a loop in which time is incremented by $\Delta t$  at each iteration; the number of iterations is chosen nondeterministically (lines \arabic{innerstart}--\arabic{innerend}).

Using this model, we are able to find the same defect reported in the earlier work.  This manifests as a violation of the assertion $K>0$ (line \arabic{assertK}) found with $N=2$ and $M=1$ in less one second.  The model checker also prints the symbolic values of all variables that lead to the violation.  When function \verb!computeK! is replaced with the corrected version, no violations are found.  For $N=M=3$, verification takes $1.5$ seconds.  This time includes 31 calls to the automated theorem prover Z3 \cite{demoura-bjorner:2008:z3}, which easily discharges the assertions.
The CIVL approach provides strong evidence for the correctness of the fix, though not as strong as a formal proof.  Nevertheless, it was straightforward to write the model and the verification process itself is fully automated.

Although tools like KeYmaera~X provide the additional rigor of formal proofs, lightweight tools such as CIVL can play a critical role in the early stages of the design and implementation process. The ability to quickly find defects and receive automated feedback helps developers iterate rapidly and refine their models without the overhead of constructing formal proofs at each step. Once these preliminary checks confirm the absence of obvious errors, investing additional effort in proof tools like KeYmaera~X becomes more strategic and meaningful.


Finally, CIVL’s support for modeling concurrency, a fundamental feature of many scientific codes, highlights a complementary strength compared to tools like KeYmaera~X, which focus on formal proofs but lack concurrency support. While concurrency was not required for this specific problem, this capability, combined with CIVL’s lightweight and automated approach, underscores the broader potential for lightweight formal methods to assist developers of ocean models in debugging and analyzing their algorithms effectively.

\section{Conclusions}

Formal methods offer a promising path toward improving the reliability and trustworthiness of ESMs. Given the complexity of ESMs and their reliance on bitwise reproducibility and manual analysis, formal specification and verification can serve as a valuable complementary approach. By defining explicit correctness properties and systematically checking them, formal methods can bolster current practices, reduce subjectivity in model assessments, and help detect subtle software or algorithmic defects that might otherwise go unnoticed.

Lightweight formal methods, in particular, like those employed by CIVL offer a practical and scalable approach to verifying numerical software. CIVL’s syntax and semantics closely resemble imperative programming languages, making it accessible to scientific software developers while enabling automated verification. Ability to rapidly detect defects and provide automated feedback allows developers to iterate efficiently without the overhead of constructing full formal proofs at every step. Our case study, for instance, demonstrated how CIVL can uncover subtle defects in ocean model parameterization schemes and validate corrections efficiently. Expanding the use of these lightweight verification techniques in ESMs, particularly for parallelism, floating-point precision, and complex data dependencies, could significantly enhance model reliability.

Our work highlights several essential features of ESMs that are well suited for formal reasoning, including component coupling and concurrency, the interplay of continuous and discrete processes, and the intricate spatial discretization considerations. We have proposed abstraction approaches for these complexities to allow for more systematic verification of model components. Although our initial exploration demonstrates the feasibility of using formal methods, we recognize that substantial effort is needed to create tools and abstractions that climate scientists can readily adopt.

A crucial next step involves addressing correctness in large-scale GPU simulations, such as those implemented with AMReX, A software framework for massively parallel, block-structured AMR applications~\cite{zhang2019amrex}. While AMReX provides essential capabilities like domain periodicity, implementing specialized boundary handling, such as for tripolar grids, introduces significant concurrency and correctness challenges. Formal specification can help ensure that new routines, parameterization logic, and hardware adaptations maintain critical invariants, such as stable boundary operations, through automated verification. Moreover, formal methods can proactively address potential data races in concurrent execution and mitigate numerical inconsistencies that may arise from compiler variations or hardware differences~\cite{cindy-dolores-compiler-induced-inconsistencies-ics-2024}, improving the robustness and reproducibility of large-scale climate simulations.

Moving forward, bridging the gap between formal methods research and climate modeling practice will require interdisciplinary collaboration. By developing tools that integrate seamlessly into existing workflows, we can enable broader adoption of specification and verification techniques. Our long-term vision is to establish formal methods as a complementary component of climate model development, enhancing both scientific rigor and predictive reliability.

\subsection*{Acknowledgments.}  S.F.\ Siegel was supported by U.S.\ National Science Foundation grant CCF-1955852 and by the  RAPIDS Institute, U.S.\ Department of Energy, Office of Science, Office of Advanced Scientific Computing Research, Scientific Discovery through Advanced Computing (SciDAC) program, under Award Number DE-SC0021162.
G.\ Gopalakrishnan was supported by U.S.\ National Science Foundation grants CCF 2446084, 2403379, 2346394 and 2217154. J.\ Baugh was supported by U.S.\ National Science Foundation grant  CCF 2124205.

This material is based upon work supported by the U.S. National Science Foundation National Center for Atmospheric Research, which is a major facility sponsored by the U.S. National Science Foundation under Cooperative Agreement No. 1755088. Any opinions, findings, and conclusions or recommendations expressed in this material are those of the author(s) and do not necessarily reflect the views of the U.S. National Science Foundation.

This report was prepared as an account of work sponsored by an agency of the United States Government. Neither the United States Government nor any agency thereof, nor any of their employees, makes any warranty, express or implied, or assumes any legal liability or responsibility for the accuracy, completeness, or usefulness of any information, apparatus, product, or process disclosed, or represents that its use would not infringe privately owned rights. Reference herein to any specific commercial product, process, or service by trade name, trademark, manufacturer, or otherwise does not necessarily constitute or imply its endorsement, recommendation, or favoring by the United States Government or any agency thereof. The views and opinions of authors expressed herein do not necessarily state or reflect those of the United States Government or any agency thereof.

\nocite{*}
\bibliographystyle{eptcs}
\bibliography{generic}
\end{document}